\documentclass[pra,showpacs,twocolumn,superscriptaddress]{revtex4}
\usepackage[utf8]{inputenc}
\usepackage[american,british]{babel}
\usepackage[T1]{fontenc}
\usepackage[pdftex]{graphicx}  
\usepackage{graphicx}
\usepackage{dcolumn}
\usepackage{bm}
\usepackage{amsmath,amsthm,amssymb}
\usepackage{color}
\usepackage{epsfig}
\newcommand{\bra}[1]{\left\langle #1 \right|}
\newcommand{\ket}[1]{\left| #1 \right\rangle}

\newcommand{\ketbra}[2]{\left|#1\middle\rangle\middle\langle#2\right|}

\newcommand{\norm}[1]{\left\|#1\right\|}

\newcommand{\M}[1]{\mathcal{#1}}
\newcommand{\Md}[2]{\mathcal{#1}^{#2}}

\newcommand{\bd}[1]{b^{\dagger}_{#1}}
\newcommand{\ad}[1]{a^{\dagger}_{#1}}

\newcommand{\vacket}{\ket{vac}}
\newcommand{\vacbra}{\bra{vac}}

\input{Qcircuit}

\begin{document}
\title{Quantumness of correlations in indistinguishable particles}
\author{Fernando Iemini}
\email{fernandoiemini@gmail.com}

\affiliation{Departamento de F\'{\i}sica - ICEx - Universidade Federal de Minas Gerais,
Av. Pres.  Ant\^onio Carlos 6627 - Belo Horizonte - MG - Brazil - 31270-901.}

\author{Tiago Debarba} 
\affiliation{Departamento de F\'{\i}sica - ICEx - Universidade Federal de Minas Gerais,
Av. Pres.  Ant\^onio Carlos 6627 - Belo Horizonte - MG - Brazil - 31270-901.}
\affiliation{Institute for Quantum Computing,
University of Waterloo, Waterloo ON N2L 3G1, Canada}

\author{Reinaldo O. Vianna}
\email{reinaldo@fisica.ufmg.br}
\affiliation{Departamento de F\'{\i}sica - ICEx - Universidade Federal de Minas Gerais,
Av. Pres.  Ant\^onio Carlos 6627 - Belo Horizonte - MG - Brazil - 31270-901.}
\date{\today}

\begin{abstract} 
We discuss a general notion of quantum correlations in
 fermionic or bosonic indistinguishable particles. Our
 approach is mainly based on the identification of the algebra
 of single-particle observables, which allows us to devise an
 activation protocol in which the \textit{quantumness of correlations} in the
 system leads to a unavoidable creation of entanglement with
 the measurement apparatus. Using the distillable entanglement,
 or the relative entropy of entanglement, as entanglement measure,
 we show that our approach is equivalent to the notion of
 minimal disturbance in a single-particle von Neumann measurement,
 also leading to a geometrical approach for its quantification.

\end{abstract}

\maketitle

\section{Introduction}

The notion of entanglement, first noted by Einstein,
Podolsky, and Rosen \cite{epr35}, is considered one of the main
features of quantum mechanics, and became a subject of
great interest in the last few years due to its primordial
role in quantum computation and quantum information
 \cite{Nielsen,Horodecki09,Vedral08,Kais}.
 However, entanglement is not the only kind of correlation
 presenting non classical features, and a great effort has recently
 been directed towards characterizing a more general notion
 of quantum correlations, the \textit{quantumness} of correlations.
 The quantumness of correlations is revealed in different ways,
 and there are a wide variety of approaches, sometimes equivalent,
 in order to characterize and quantify it, e.g., through the  
 ``activation protocol'', where the non classical character
 of correlations in the
 system is revealed by a unavoidable creation of entanglement
 between system and measurement apparatus in a local
 measurement \cite{streltsov11,piani11};
 or by the analysis of the minimum disturbance caused in the system by
 local measurements \cite{zurek01,vedral01,luo08},
 which led to the seminal definition
 of \textit{quantum discord} \cite{zurek01};
 or even through geometrical approaches \cite{modi10}.

Despite being widely studied in systems of distinguishable particles, less attention
has been given to the study of entanglement,
 or even a more general notion of quantum correlations,
 in the case of indistinguishable particles. In this case, the space of quantum
states is restricted to symmetric $\M{S}$ or antisymmetric $\M{A}$
subspaces, depending on the bosonic or fermionic nature of the
system, and the particles are no longer accessible individually,
 thus eliminating the usual notions of separability and local measurements, and
 making the analysis of correlations much subtler.
In fact, there are a multitude of distinct approaches and
 an ongoing debate around the entanglement
 in these systems \cite{balachandran,ghirardi,
schliemann01a01b,eckert02,li01,zanardi02,wiseman03,benatti10,banulus07,
grabowski1112,tsutsui}. Nevertheless, despite the variety,
 the approaches consist essentially in the
analysis of correlations under two different aspects: the
correlations genuinely arising from the entanglement between
the particles (“entanglement
of particles”) \cite{balachandran,ghirardi,
schliemann01a01b,eckert02,li01},
 and the correlations arising from the
entanglement between the modes of the system (“entanglement
of modes”) \cite{zanardi02,wiseman03,benatti10,banulus07}.
 These two notions of entanglement are
complementary, and the use of one or the other depends
on the particular situation under scrutiny. For example,
the correlations in eigenstates of a many-body Hamiltonian
could be more naturally described by particle entanglement,
whereas certain quantum information protocols could prompt
a description in terms of entanglement of modes.
 The modes notion associates a Fock space to the several
 distinguishable modes of a system of indistinguishable
particles, which allows one to employ all the tools commonly
 used in distinguishable quantum systems. The entanglement of particles
 has different definitions which agree
 in some respects, and differ in others; but once one has opted
 for a certain definition, there are also several proposed methods
 to calculate it \cite{iemini13b,iemini13a,paskauskas01,plastino09,zander10}.

Note that the correlations between  modes in a system of indistinguishable particles
 is subsumed  in the usual analysis of correlations in systems of distinguishable ones.
 Thus we shall, in this work, characterize and quantify a
 general notion of quantum correlations (not necessarily entanglement)
 genuinely arising between  indistinguishable particles.
We shall call these correlations by quantumness of correlations, to distinguish
 from entanglement, and it has an interpretation analogous
 to the quantumness of correlations in systems of distinguishable particles,
 as we shall see. One must however be careful with such phraseology,
 since systems of indistinguishable particles
 always have {\em exchange correlations} coming from the symmetric or antisymmetric nature 
 of the wavefunction. The intrinsic exchange correlations are not included in the concept of
 the quantumness of correlations.
 We shall discuss  these issues in more detail throughout the article.

The article is organized as follow. In Sec.\ref{quantumness.disting}
 we briefly review the notion of quantumness of correlations
 in distinguishable subsystems, and their interplay with the
 measurement process via the activation protocol.
 In Sec. \ref{act.prot.indist} we introduce the activation protocol
 for systems of indistinguishable particles; and in Sec.\ref{results} we characterize
 and quantify the quantumness of correlations in these systems.
We conclude in Sec.\ref{conclusion}.

\section{Quantumness of correlations}
\label{quantumness.disting}

The concept of quantumness of correlations
 is related to the amount of inaccessible information of a composed system
 if we restrict to the application
 of local measurements on the subsystems \cite{zurek01,vedral01,praopp,prlopp}. 
Since quantumness of 
correlations can be created with local operations on the subsystems,
 it is also called as the 
quantum properties of classical correlations \cite{zurek01,praopp}.
A model for the description of a measurement process is given via 
decoherence \cite{zurekreview}, 
where in order to measure a quantum system we must interact it with 
a measurement apparatus, which is initially uncorrelated with the quantum system.
 This interaction, given by a unitary evolution, creates correlations
 between them, and thereby the measurement outcomes
 will be registered on the apparatus eigenbasis.
 A protocol that allows us to understand 
the interplay between a measurement process and 
the quantumness of correlations in a system
is known as the {\it nonclassical correlations activation protocol}.
This protocol shows that if and only if the system is strictly classically correlated,
 i.e., has no quantumness of correlations, 
 there exists a local measurement on the subsystems that does not
 create entanglement between system and measurement apparatus
 \cite{streltsov11, piani11, gharibianijqi}; or rather,
 if the system has quantumness of correlations,
 then it will  inevitably create entanglement with the
 apparatus measurement in a local measurement process,
 hence the reference to ``activation''. A direct corollary of this protocol allows us to quantify the amount 
of quantumness of correlations by measuring the minimal amount 
of entanglement created 
between the system and the measurement apparatus during a local measurement 
process \cite{pianiadessorapcom}.

 Given, for instance, a bipartite system $S$ 
described by the 
state $\rho_S$, in order to apply a von Neumann measurement in this system we 
must interact it with a measurement apparatus $\M{M}$, initially in an arbitrary state
 $\ketbra{0}{0}_\M{M}$. Suppose that we are able to apply global von Neumann measurements 
in such a system, e.g., a von Neumann measurement in the system eigenbasis $\{\ket{i}\}_{\M{S}}$,
 $\rho_S=\sum_i\lambda_i\ketbra{i}{i}$.
 The  system and the measurement apparatus must interact under the action 
of the following unitary transformation: 
$U_{S:\M{M}}\ket{i}_S\ket{0}_{\M{M}}=\ket{i}_S\ket{i}_\M{M}$.
We see that the interaction simply 
creates classical correlations between them: 
$\tilde{\rho}_{S:A} = U_{S:A}( \rho_S\otimes\ketbra{0}{0}_A )U_{S:A}^{\dagger}
=\sum_i\lambda_i\ketbra{i}{i}\otimes\ketbra{i}{i}$. 
If however we are restricted to apply local measurements, the measurement 
process will create entanglement between system and 
apparatus by their corresponding coupling unitary $U'_{S:A}$, unless the 
state is strictly classically correlated, as stated by the activation protocol.
  The minimal amount of entanglement 
$E(\tilde{\rho}_{S:A})$ which is 
created in a local measurement process is quantified by the quantumness 
of correlations $Q(\rho_S)$ of the system, i.e.,
\begin{equation}
Q(\rho_S) = \min_{U_{S:A}}E(\tilde{\rho}_{S:A}).
\end{equation} 

Different entanglement
measures will lead, in principle, to different quantifiers
for the quantumness of correlations. The only requirement
 is that the entanglement measure be monotone
 under LOCC maps \cite{piani11,streltsov11,pianiadessorapcom}. 
Other measures of quantumness can be recovered with the activation protocol: 
the quantum discord \cite{streltsov11}, one-way work deficit \cite{streltsov11}, 
relative entropy of quantumness \cite{piani11} and the geometrical measure 
of discord via trace norm \cite{taka}, are some examples.

\section{Activation protocol for indistinguishable particles}
\label{act.prot.indist}
As aforesaid, quantum correlations between distinguishable
 particles can be interpreted via a unavoidable entanglement
 created with the measurement apparatus in a partial von Neumann
 measurement on the particles \cite{streltsov11,piani11}, i.e.,
 in a measurement corresponding to a non-degenerate local observable.
Note that although the approach is
based on projective measurements, it is valid and well defined also
for POVMs: once the dimension and the partitioning 
of the ancilla can be arbitrarily chosen, general measurements 
can be realized through a direct application of the Naimark's dilation theorem.
 In systems of indistinguishable particles the notion
 of ``local measurement'' will be implemented through the
 algebra of single-particle observables (see for
 example Ref.\cite{balachandran} for a detailed discussion),
 and based on this identification we shall set up an ``activation
 protocol'' for indistinguishable particles. The importance to study the correlations,
 particularly the entanglement, in terms of subalgebras of observables
 has been emphasized in \cite{balachandran,zanardi0104,
benatti10,barnum04,harshman11,derkacz12},
 proving to be a useful approach for such analysis.
 The algebra of single-particle observables
 is generated by,
\begin{eqnarray}
\M{O}_{sp} = &M \otimes \M{I} \otimes \cdots \otimes \M{I} +
 \M{I} \otimes M \otimes \cdots \otimes \M{I} + \cdots + \nonumber\\
& \M{I} \otimes  \cdots \otimes \M{I} \otimes M, 
\end{eqnarray}
where $M$ is an observable in the Hilbert space of a single particle.
 We can express this algebra in terms
 of fermionic or bosonic creation $\{\ad{i}\}$ and annihilation $\{a_i\}$
 operators, depending on the nature of the particles in the system.
 The algebra is generated by quadratic observables
 $\M{O}_{sp} = \sum_{ij} M_{ij}\ad{i}a_j$ that can be diagonalized
 as $\M{O}_{sp} = \sum_k \lambda_k \tilde{a}^{\dagger}_k \tilde{a}_k$,
 where $\tilde{a}^{\dagger}_k = \sum_j U_{kj} \ad{j}$ and $U$ is
 the unitary matrix which diagonalizes $M$.
 Thus, since it is a non-degenerate algebra,
 the eigenvectors of their single-particle observables will be given
 by single Slater determinants, or permanents, for fermionic and
 bosonic particles respectively; more precisely, given by the set
 $\{\tilde{a}^{\dagger}_{\vec{k}}\vacket\}$ where $\vec{k}
 = (k_1,\cdots,k_n)$,
 $k_i \in \{1,2,...,dim_{single-particle}\}$, represents the states
 of occupation of $n$ particles, $\tilde{a}^{\dagger}_{\vec{k}} =
 \tilde{a}^{\dagger}_{k_1}\tilde{a}^{\dagger}_{k_2}
\cdots\tilde{a}^{\dagger}_{k_n}\ket{vac}$,
 $dim_{single-particle}$ is the single-particle dimension 
 and $\vacket$ is the vacuum state. The measurement of
 single-particle observables is therefore given by a
 von Neumann measurement, which we shall call hereafter as single-particle
 von Neumann measurement, according to the complete set of rank one
 projectors $\{\tilde{\Pi}_{\vec{k}} =
 \tilde{a}^{\dagger}_{\vec{k}}\vacket\vacbra  \tilde{a}_{\vec{k}}\}$,
 $\sum_{\vec{k}} \tilde{\Pi}_{\vec{k}} = \M{I}_{\M{A}(\M{S})}$, being $\M{I}_\M{A}$
 and $\M{I}_\M{S}$ the identity of the antisymmetric and
 symmetric subspaces, respectively. 

Recall that a measurement
 can be described by coupling the system to a measurement
 apparatus, being the measurement outcomes obtained by measuring the apparatus
 in its eigenbasis. Given a quantum state $\rho_{\M{Q}}$, and
 a measurement apparatus $\M{M}$ in a pure initial state $\ket{0}_{\M{M}}$, such that
 $\rho_{\M{Q},\M{M}} = \rho_{\M{Q}} \otimes \ketbra{0}{0}_{\M{M}}$,
their coupling is given by applying a unitary $U$ on the total state
 that will correlate system and apparatus,
$\tilde{\rho}_{\M{Q},\M{M}} = U (\rho_{\M{Q}}
 \otimes \ketbra{0}{0}_{\M{M}}) U^{\dagger}$.
 Such unitary $U$ realizes a single-particle
 von Neumann measurement $\{\Pi_{\vec{k}}\}$ 
if for any quantum state $\rho_{\M{Q}}$ holds:
 $Tr_{\M{M}}(U (\rho_{\M{Q}} \otimes \ketbra{0}{0}_{\M{M}}) U^{\dagger}) = 
\sum_{\vec{k}} \Pi_{\vec{k}} \rho_{\M{Q}} \Pi_{\vec{k}}^{\dagger}$.

It is not hard to see how the unitary $U$ must act in order to realize the
$\{\Pi_{\vec{k}} = \ad{\vec{k}}\vacket\vacbra a_{\vec{k}}\}$,
 $\sum_{\vec{k}} \Pi_{\vec{k}} = \M{I}_{\M{A}(\M{S})}$ measurement.
 Let us first consider
 the following notation, $\{\ad{\vec{k}}\ket{vac}\} =
 \left\{\ket{f(\vec{k})}\right\}$, $f(\vec{k}) \in \{1,2,..,dim_{\M{A}(\M{S})}\}$,
 being $f$ a bijective function of the sets $\{\vec{k}\}$ and
 $\{1,2,..,dim_{\M{A}(\M{S})}\}$, and $dim_{\M{A}(\M{S})}$ is the dimension of
 the antisymmetric or symmetric subspaces.
 Given that the apparatus has at least the same dimension
 as the system, the unitary is given by,
\begin{equation}
U \ket{f(\vec{k})}_{\M{Q}} \otimes \ket{j}_{\M{M}} =
 \ket{f(\vec{k})}_{\M{Q}} \otimes \ket{j \oplus f(\vec{k})}_{\M{M}}.
\label{coupling.unitary}
\end{equation}
It is easy to show that such operator is indeed unitary;
 note that
\begin{equation}
U = \sum_{\vec{k},j} \ket{f(\vec{k})}
\ket{j\oplus f(\vec{k})}\bra{f(\vec{k})}\bra{j},
\end{equation}
 thus, 
 {\small  \begin{equation}
U U^{\dagger} = \sum_{\vec{k},j,\vec{k'},j'}
 \delta_{\vec{k},\vec{k'}}\delta_{j,j'}
 \ket{f(\vec{k})}\ket{j\oplus f(\vec{k})}
\bra{f(\vec{k'})}\bra{j'\oplus f(\vec{k'})},
\end{equation}}
 and since $\{\ket{f(\vec{k})}_{\M{Q}}\}_{\vec{k}}$
 and $\{\ket{j \oplus f(\vec{k})}_{\M{M}}\}_j$
 form a complete set,
 we have that $U U^{\dagger} = \M{I}_{\M{A}(\M{S})} \otimes \M{I}_{\M{M}}$.

Defined the coupling unitary, we are now
 able to analyze the entanglement created between system
 and apparatus in a single-particle von Neumann measurement, $E_{\M{Q},\M{M}}$.
 Given a quantum state $\rho_{\M{Q}}$,
 we intend to quantify the minimum of such entanglement over all
 single-particle von Neumann measurements,
 $\min_U E_{\M{Q},\M{M}}\left[U (\rho_{\M{Q}}
 \otimes \ketbra{0}{0}_{\M{M}}) U^{\dagger}\right]$.
This quantity then corresponds to the quantumness of correlation
 in systems of indistinguishable particles.
 Note that such minimization is analogous to the activation protocol  given in \cite{piani11},
 but now for systems of indistinguishable particles, where the
 minimization is carried out on the single-particle
 unitary transformations $V^{\otimes n}$, see Fig.\ref{activation.protocol.fig}.

\begin{center}
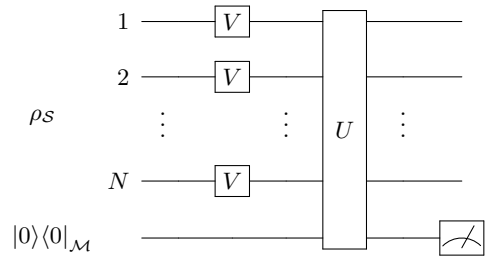
\begin{figure}[t]
\mbox{
\Qcircuit @C=1.5em @R=1.0em {  
& & & \lstick{1}           & \qw & \gate{V} & \qw &   \multigate{5}{U} &\qw    & \qw \\
& & & \lstick{2}           & \qw & \gate{V} & \qw &          \ghost{U} & \qw   &\qw \\
&\lstick{\rho_{\mathcal{S}}}& &          &  \lstick{\vdots}&  &\vdots &  &  \vdots &   &   &\\  
& & &                      &     &          &           &      &  &\\
& & & \lstick{N}           & \qw & \gate{V} & \qw   &        \ghost{U} & \qw   & \qw \\
& & \lstick{\ketbra{0}{0}_{\mathcal{M}}}& &\qw & \qw & \qw & \ghost{U} & \qw   & \meter } } 
\caption{Activation protocol
 for a system of indistinguishable particles, where $\rho_\mathcal{S}$ 
 is the state of the system, $\ketbra{0}{0}_{\mathcal{M}}$ 
 represents the measurement apparatus, $V$ is
 the single-particle unitary transformations and $U$ the
 unitary (as given by Eq.(\ref{coupling.unitary}))
 respective to a single-particle von Neumann measurement.}
\label{activation.protocol.fig}
\end{figure}
\end{center}
\section{Results}
\label{results}
Regardless of which entanglement measure is used, let us first
 see which set of states does not generate entanglement
 after the activation protocol, i.e., has no quantumness of correlations.
 We find that
 this set $\{\xi\}$ is given by states that
 possess a convex decomposition in orthonormal
 pure states described by single Slater determinants, or permanents,
\begin{equation}
\xi = \sum_{\vec{k}} p_{\vec{k}}\,
 \tilde{a}^{\dagger}_{\vec{k}}\vacket\vacbra \tilde{a}_{\vec{k}},
 \quad \sum_{\vec{k}} p_{\vec{k}} = 1,
\label{state.with.no.part.correl}
\end{equation}
where $\tilde{a}^{\dagger}_{\vec{k}}\vacket =
 V^{\otimes n} \ad{\vec{k}}\vacket$, $V$ is a unitary matrix,
 and $\{\ad{\vec{k}}\}$ an orthonormal set of creation operators.

\begin{proof} We shall first show that states given
 by Eq.(\ref{state.with.no.part.correl}) do
 not generate entanglement, and then that they are the only ones.
 Let $U$ be the coupling unitary corresponding to the
$\{\Pi_{\vec{k}} = \ad{\vec{k}}\vacket\vacbra a_{\vec{k}} =
 \ket{f(\vec{k})}\bra{f(\vec{k})}\}$,
 $\sum_{\vec{k}} \Pi_{\vec{k}} = \M{I}_{\M{A}(\M{S})}$ measurement. 
 Applying the activation protocol on states given by Eq.(\ref{state.with.no.part.correl}),
 using $\bar{V} = V^{\dagger}$ as the single-particle unitary
 transformation, it follows that: 
\begin{eqnarray}
\rho_{\M{Q}:\M{M}} &=& U\left[(\bar{V}^{\otimes n} \xi \bar{V}^{\dagger^{\otimes n}} )_{\M{Q}}
\otimes \ketbra{0}{0}_{\M{M}} \right]U^{\dagger} \\
&=& \sum_{\vec{k}} \,p_{\vec{k}}\,
\ketbra{f(\vec{k})}{f(\vec{k})}_{\M{Q}} \otimes
\ketbra{f(\vec{k})}{f(\vec{k})}_{\M{M}} \nonumber, 
\end{eqnarray}
where $\rho_{\M{Q}:\M{M}}\in Sep(\M{Q}\otimes\M{M}).$
 The demonstration
 that such states correspond to the unique states that do not
 generate entanglement is given below. A separable state
 between system and measurement apparatus can be given by,
\begin{equation}
\sigma = \sum_i p_i \ketbra{\psi_i}{\psi_i}_{\M{Q}}\otimes
 \ketbra{\phi_i}{\phi_i}_{\M{M}},
\end{equation}
noting that the sets $\{\ket{\psi_i}\}$ and $\{\ket{\phi_i}\}$
 are not necessarily orthogonal.
Since the activation protocol corresponds to a unitary operation,
 thus invertible, there must exist a set $\{\ket{\eta_i}\}$ of
 states for the system such that,
\begin{equation}
U (V^{\otimes n})\, \ket{\eta_i}_{\M{Q}}\otimes \ket{0}_{\M{M}}
 = \ket{\psi_i}_{\M{Q}}\otimes \ket{\phi_i}_{\M{M}},
\label{sup.1}
\end{equation}
and $\rho_{\M{Q}} = \sum_i \, p_i \, \ket{\eta_i}\bra{\eta_i}$.
 Expanding $\{\ket{\eta_i}\}$ on the basis $\{\ad{\vec{k}}\vacket\}$
 ``transformed'' by $V^{\dagger^{\otimes n}}$,
\begin{equation}
\ket{\eta_i} = \sum_{\vec{k}} c_{\vec{k}}^{(i)} \,
 V^{\dagger^{\otimes n}}\,\ad{\vec{k}}\vacket,
\label{sup.2}
\end{equation}
we see from Eqs.(\ref{sup.1}) and (\ref{sup.2}) that,
\begin{equation}
U (V^{\otimes n})\, \ket{\eta_i}\otimes \ket{0} =
 \sum_{\vec{k}} c_{\vec{k}}^{(i)} \,\ad{\vec{k}}
 \vacket \otimes \ket{f(\vec{k})} = \ket{\psi_i}\otimes \ket{\phi_i}.
\end{equation}

The above factorization condition imposes the following
 restriction:
  $ c_{\vec{k}}^{(i)} = \gamma_i\, \delta_{ \{ \vec{k},g(i) \} }, \,\norm{\gamma_i} =
 1,\,g : \{i\} \mapsto \{ \vec{k} \} $. Therefore,
 \begin{eqnarray} 
\rho_{\M{Q}} &=& \sum_i p_i \,\ket{\eta_i}\bra{\eta_i},\nonumber \\
&=& \sum_i p_i \, (\sum_{\vec{k}} \gamma_i \,
\delta_{\{\vec{k},g(i)\}}\, \ad{\vec{k}} \vacket) \\
& & 
 (\sum_{\vec{k'}} \vacbra a_{\vec{k'}}\,\gamma_i^*\,
 \delta_{\{\vec{k'},g(i)\}}), \nonumber\\
&=& \sum_i p_i \,\underbrace{\norm{\gamma_i}}_{1}\,
\ad{g(i)}\vacket \vacbra a_{g(i)}, \nonumber 
\end{eqnarray}  i.e, the states with no quantumness of correlations as given by Eq.(\ref{state.with.no.part.correl}).
\end{proof}

\textit{Example.} Let us show an example of the
 approach in order to clarify the formalism and the above analysis.
 An interesting case concerns to the controversial bosonic quantum
 state $\ket{\psi_b} = \frac{1}{2}(\bd{0}\bd{0} + \bd{1}\bd{1})\vacket
 \in \M{S}(\Md{H}{2}\otimes\Md{H}{2})$, where $\{\bd{i}\}$ are the
 bosonic creation operators. Such a state is considered
 both entangled by some authors \cite{eckert02,grabowski1112,paskauskas01},
 as non entangled for others \cite{balachandran,ghirardi,li01}.
 Note that such a state can actually be described by a single Slater permanent
 $\ket{\psi_b} = \bd{+}\bd{-} \vacket$, being $\bd{\pm} =
 \frac{1}{\sqrt{2}}(\bd{0} \pm \bd{1})$. 
Defining the coupling unitary $U$ corresponding to the $\{\Pi_{\vec{k}} =
 \bd{\vec{k}}\vacket\vacbra b_{\vec{k}}\}$,
 $\sum_{\vec{k}} \Pi_{\vec{k}} = \M{I}_{\M{S}}$, $\{\vec{k}\} =
 \{(0,0),(0,1),(1,1)\}$ measurement, and using the notation,
\begin{eqnarray}
\bd{0}\bd{0} \vacket = \ket{0},\, \bd{0}\bd{1} \vacket = \ket{1},
\, \bd{1}\bd{1} \vacket = \ket{2},
\end{eqnarray}
we have that the unitary acts as follows,
\begin{equation}
U \ket{k}_{\M{Q}}\otimes \ket{0}_{\M{M}} = \ket{k}_{\M{Q}}\otimes \ket{k}_{\M{M}}.
\end{equation}
Applying this unitary on the bosonic state, we generate an entangled
 state between system and apparatus, $U (\ket{\psi_b}_{\M{Q}}\otimes
 \ket{0}_{\M{M}}) =  \frac{1}{2}(\bd{0}\bd{0}\vacket\otimes \ket{0}
 + \bd{1}\bd{1}\vacket\otimes \ket{2})$, but this is not a unavoidable
 entanglement in order to realize that measurement, since we could apply,
 before the unitary coupling, the following single-particle unitary
 transformation, $V: \ket{+} = \ket{0} + i\ket{1} \mapsto \ket{0},
 \ket{-} = \ket{0} - i\ket{1} \mapsto \ket{1}$, i.e,
\begin{eqnarray}
V\otimes V: \left\{ \begin{array}{l}
\bd{+} \mapsto \bd{0},\\
\bd{-} \mapsto \bd{1}.
\end{array}\right.
\end{eqnarray}
We see now that the coupling between system and apparatus does not
 generate entanglement between them,
 $U \left[(V\otimes V)\ket{\psi_b}_{\M{Q}}\otimes \ket{0}_{\M{M}}\right] =
  U(\bd{0}\bd{1}\vacket_{\M{Q}} \otimes \ket{0}_{\M{M}}) =
 \bd{0}\bd{1}\vacket_{\M{Q}} \otimes \ket{1}_{\M{M}} \in
 Sep(\M{Q}\otimes\M{M})$, and thus such a state has no quantumness of correlations.


An important result to be emphasized in this analysis
 via the activation protocol relates to
 the establishment of an equivalence between the quantumness of
 correlations with the \textit{distinguishable} bipartite entanglement
 between system and apparatus, showing the
 usefulness of the correlations between indistinguishable particles.
Note that the set $\{\xi\}$ is simply the antisymmetrization or
 symmetrization of the distinguishable classically correlated
 states (states with distinguishable particles with no
 quantumness of correlations), and all their correlations
 are due to the exchange correlations; the activation protocol then shows that
any kind of correlations between indistinguishable particles
 beyond the mere exchange correlations can always be activated or mapped
 into distinguishable bipartite entanglement between $\M{Q}:\M{M}$.

\begin{figure}
\includegraphics[scale=0.5]{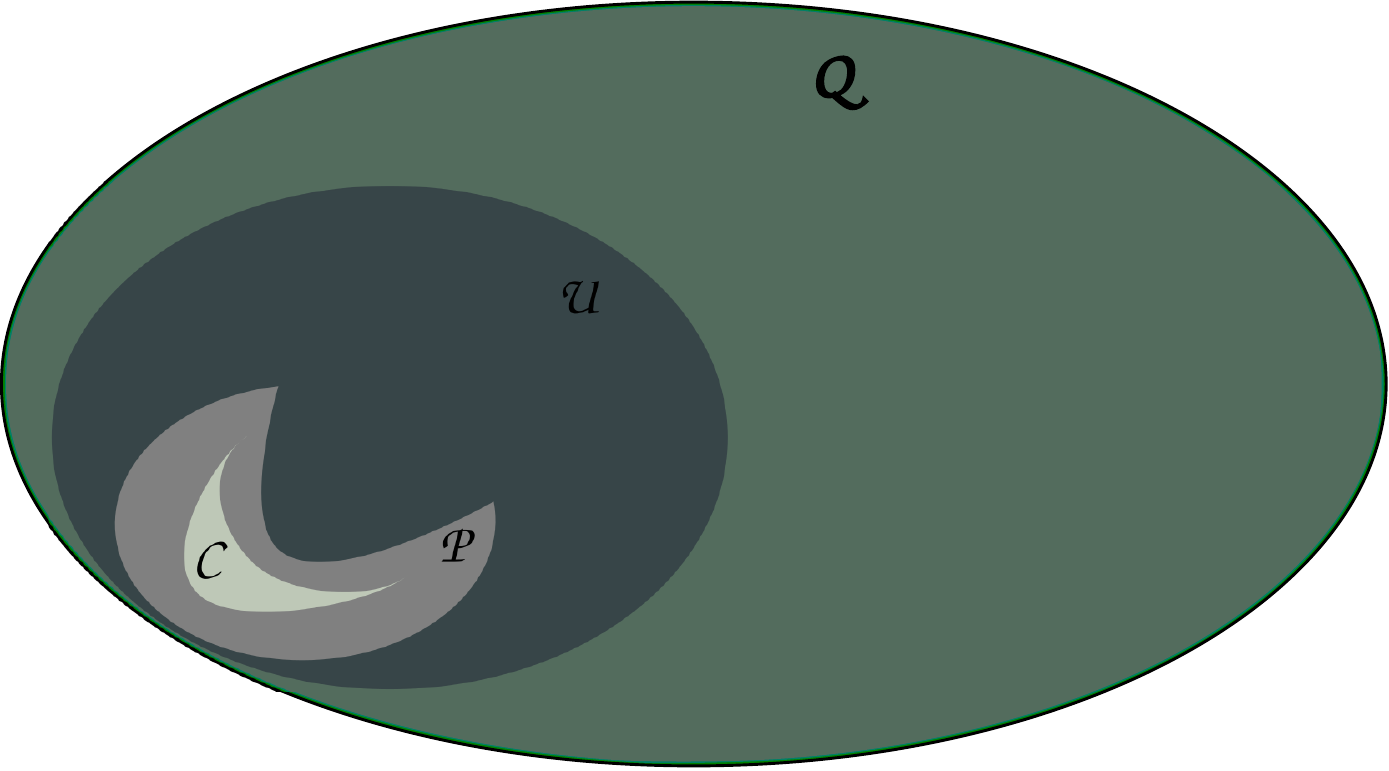}
\caption{(Color online) Schematic picture of the distinct types of correlations
 in systems of indistinguishable particles. The larger set ($\M{Q}$) denotes the set
 of all fermionic, or bosonic, quantum states; the blue area ($\M{U}$)
 represents the convex set of states with no entanglement; the gray area ($\M{P}$)
 represents the non convex set of states with no quantumness of correlations,
 as defined in this article (Eq.(\ref{state.with.no.part.correl}));
 and the yellow area ($\M{C}$) represents the non convex set of
 states with no exchange correlations due to the particle statistics,
 possessing only classical correlations. Note that for
 fermionic particles, the set $\M{C}$ is a null set.
 The following hierarchy is identified:
 $\M{C} \subset \M{P} \subset \M{U} \subset \M{Q}$.}
\label{correlation.sets.fig}
\end{figure}

The correlations between indistinguishable particles can thereby
 be characterized by different types:
the entanglement,
 the quantumness of correlations as discussed in this article,
 the correlations generated merely by particle statistics (exchange correlation),
 and the classical correlations. 
In fact, there are quantum states
 whose particles are classically correlated, not even possessing
 exchange correlations, such as pure bosonic states with
 all their particles occupying the same degree of freedom, $\ket{\psi_b} =
 \frac{1}{\sqrt{n!}}(\bd{i})^n\vacket$,
 or mixed states described by an orthonormal convex
 decomposition of such pure states,
 $\chi_b = \sum_i \frac{1}{n}(\bd{i})^n \vacket \vacbra (b_{i})^n$.
 See Fig.\ref{correlation.sets.fig} for
 a schematic picture of these different kinds of correlations.
 Interesting questions to raise concern how the notion of 
 entanglement of particles
 is related to the quantumness of correlations,
 and if  they are equivalent for pure states.
We can note from Eq.(\ref{state.with.no.part.correl}) that, for pure states, the set with no quantumness
 of correlations is described by states with a single Slater determinant, or permanent,
 which is equivalent to the set of unentangled pure states. 
Actually there is an ongoing debate regarding
 the correct definition of particle entanglement \cite{balachandran,ghirardi,
schliemann01a01b,eckert02,li01}, but at the same time there
 are strong physical reasons to consider particle entanglement in pure states as the 
 correlations beyond the mere exchange correlations
 \cite{balachandran,ghirardi,schliemann01a01b,li01}. 
 Concerning  mixed states,
 it becomes clear that the set given by Eq.(\ref{state.with.no.part.correl})
 is a subset of the unentangled one, thereby being quantumness of correlations
 a more general notion of correlations than entanglement.

 According to the activation protocol, different entanglement measures
 will  lead, in principle, to different quantifiers for the quantumness of correlations.
 We can thus define the measure $Q_E$ for quantumness of correlations,
 associated with the entanglement measure $E$, as follows,
\begin{equation}
Q_E(\rho_{\M{Q}}) = \min_V E(\tilde{\rho}_{\M{Q},\M{M}}),
\label{activ.prot}
\end{equation}
where $\tilde{\rho}_{\M{Q},\M{M}} =
U\left[(V^{\otimes n} \rho_{\M{Q}}
 V^{\dagger^{\otimes n}} )\otimes \ketbra{0}{0}_{\M{M}}
\right]U^{\dagger}$.

We shall consider two different entanglement measures
 for the bipartite entanglement,
 the physically motivated distillable
 entanglement $E_D$ \cite{destent} 
 and the relative entropy of entanglement $E_r$
 \cite{VPqree,quantifyingent}.
 Note that the output states of the
 activation protocol have  the so called maximally correlated form \cite{rains01}
 between system and measurement apparatus,
 $\tilde{\rho}_{\M{Q},\M{M}} = \sum_{\vec{l},\vec{l'}} \,\chi_{\vec{l},\vec{l'}}^V\,
\ketbra{f(\vec{l})}{f(\vec{l'})}_{\M{Q}} \otimes 
\ketbra{f(\vec{l})}{f(\vec{l'})}_{\M{M}}$, being
 $\chi_{\vec{l},\vec{l'}}^V = (\Pi_{\vec{l}}^V)^{\dagger}
 \rho_{\M{Q}} (\Pi_{\vec{l'}}^V)$, where
$\Pi_{\vec{l}}^V = V^{\otimes n} \Pi_{\vec{l}}$ (see appendix).
 It is known that the entanglement
 for such states according to the distillable entanglement \cite{hiroshima04},
 as well as for the relative entropy of entanglement \cite{rains01},
 is given by $E_{D(r)} (\tilde{\rho}_{\M{Q},\M{M}}) = S(\tilde{\rho}_{\M{Q}})
 - S(\tilde{\rho}_{\M{Q},\M{M}})$, where $S(\rho) = -Tr(\rho \ln \rho)$
 is the von Neumann entropy. The first term is given by
 $S(\tilde{\rho}_{\M{Q}}) = S\left(\sum_{\vec{l}} (\Pi_{\vec{l}}^V)^{\dagger}
 \rho_{\M{Q}} (\Pi_{\vec{l}}^V) \ketbra{f(\vec{l})}{f(\vec{l})}\right)$, i.e.,
 the entropy of the projected state $\rho_\M{Q}$ according to a
 single-particle von Neumann measurement, and the second term is simply
 given by $S(\tilde{\rho}_{\M{Q},\M{M}}) = S(\rho_{\M{Q}})$,
 since it is invariant under unitary transformations.
 Thus we have that the quantumness of correlations measure is given by,
\begin{eqnarray}
Q_{E_{D(r)}}(\rho_{\M{Q}}) =& \min\limits_V
 \left[S\left(\sum_{\vec{l}} (\Pi_{\vec{l}}^V)^{\dagger} 
 \rho_{\M{Q}} (\Pi_{\vec{l}}^V) \ketbra{f(\vec{l})}
{f(\vec{l})}\right)\right. \nonumber\\
&-  S(\rho_{\M{Q}})],
\label{min.disturb}
\end{eqnarray}
which corresponds to the notion of minimum disturbance caused
 in the system by single-particle measurements. This result
 is in agreement with
 the analysis made in \cite{majtey13} for the particular case
 of two-fermion systems, and to the best of our knowledge is the only study
attempting to characterize and quantify a more general
 notion of correlations
 between indistinguishable particles. Using analogous
 arguments as those in \cite{modi10}, it is possible to prove
 the Eq.(\ref{min.disturb}) is an equivalent expression to,
\begin{equation}
Q_{E_{D(r)}}(\rho_{\M{Q}}) = \min_{\xi} S(\rho_{\M{Q}}\parallel\chi),
\label{geom.correlation}
\end{equation}
where $S(\rho\parallel\chi) = Tr(\rho \ln \rho - \rho \ln \chi)$ is the relative entropy.
 The above equation introduces a geometrical approach to the particle correlation measure.
 Notably we see that, as well as for the quantumness of correlations
 in distinguishable subsystems, the correlations between
 indistinguishable particles defined in this article has a
 variety of equivalent approaches in order to characterize and
 quantify it, as shown by the activation protocol (Eq.\ref{activ.prot}),
 minimum disturbance (Eq.\ref{min.disturb}) and geometrical
 approach(Eq.\ref{geom.correlation}).

\section{Conclusion}
\label{conclusion}
 In this work we discussed how to define a more general
 notion of correlation, called quantumness of correlations,
 in fermionic and bosonic indistinguishable particles,
 and presented equivalent ways to quantify it, addressing the notion
 of an activation protocol, the minimum disturbance in a single-particle
 von Neumann measurement, and a geometrical view for its quantification.
 An important result of our approach concerns to the equivalence of
 these correlations to the entanglement in distinguishable
 subsystems via the activation protocol, thus settling its usefulness
 for quantum information processing. It is interesting to note that
 the approach used in this work is essentially
 based on the definition of the algebra of single-particle observables,
 dealing here with the algebra of indistinguishable fermionic, or bosonic,
 single-particle observables, but we could apply the same idea for identical particles
 of general statistics,
 e.g. braid-group statistics, simply by defining the correct single-particle
 algebra of observables.

\acknowledgments 
We thank A.T. Cesário, T. Temistocles and T.O. Maciel for discussions.
 We acknowledge financial support by the Brazilian agencies FAPEMIG,
CNPq, and INCT-IQ (National Institute of Science and
Technology for Quantum Information).

\section*{Appendix: maximally correlated states}

Let us show that the output states of the
activation protocol for indistinguishable particles
 have the so called maximally correlated
 form between system and measurement apparatus.
 If $\{\ad{\vec{k}}\ket{vac}\} =
 \left\{\ket{f(\vec{k})}\right\}$ is the system basis, $U$ is
 the coupling unitary given by Eq.(\ref{coupling.unitary}), and
 $V$ is the unitary respective to the single particle transformation, we have that,
\begin{eqnarray}
V^{\otimes n}\, \ad{\vec{k}}\ket{vac}
 & = & (\sum_{l_1} v_{k_1 l_1}\ad{l_1}) \cdots
 (\sum_{l_n} v_{k_n l_n}\ad{l_n}) \vacket,\nonumber\\
& = & \sum_{\vec{l}} v_{k_1 l_1} \cdots
 v_{k_n l_n} \ket{f(\vec{l})},
\end{eqnarray}
where $v_{k_i l_j}$ are the matrix elements of V.
A general state for the system can be given as,
\begin{equation}
\rho_{\M{Q}} = \sum_{\vec{k},\vec{k'}} p_{\vec{k},\vec{k'}}
 \ketbra{f(\vec{k})}{f(\vec{k'})};
\end{equation}
thereby,
\begin{widetext}
\begin{eqnarray}
V^{\otimes n} \rho_{\M{Q}}
 V^{\dagger{\otimes n}} &=&
 \sum_{\vec{k},\vec{k'},\vec{l},\vec{l'}}
 p_{\vec{k},\vec{k'}} (v_{k_1 l_1} \cdots v_{k_n l_n})
\,(v_{k'_1 l'_1} \cdots v_{k'_n l'_n})^{\dagger}\,
\ketbra{f(\vec{l})}{f(\vec{l'})},\nonumber\\
&=& \sum_{\vec{l},\vec{l'}} \,\chi_{\vec{l},\vec{l'}}^V\,
\ketbra{f(\vec{l})}{f(\vec{l'})},
\end{eqnarray}\end{widetext}
where $\chi_{\vec{l},\vec{l'}}^V = \sum_{\vec{k},\vec{k'}}
 p_{\vec{k},\vec{k'}} (v_{k_1 l_1} \cdots v_{k_n l_n})
\,(v_{\vec{k'}_1 l'_1}\cdots v_{\vec{k'}_n l'_n})^{\dagger}$.
 The output states of the activation protocol thus have the form,
\begin{eqnarray}
\rho_{\M{Q}:\M{M}} &=& U\left[(V^{\otimes n} \rho_{\M{Q}} V^{\dagger^{\otimes n}} )
\otimes \ketbra{0}{0}_{\M{M}}
\right]U^{\dagger} \\
&=&\sum_{\vec{l},\vec{l'}} \,\chi_{\vec{l},\vec{l'}}^V\,
\ketbra{f(\vec{l})}{f(\vec{l'})}_{\M{Q}} \otimes 
\ketbra{f(\vec{l})}{f(\vec{l'})}_{\M{M}},\nonumber
\end{eqnarray}
i.e., the maximally correlated form.

\end{document}